\documentclass{PoS}

\usepackage{amsmath}
\usepackage{amsfonts}
\usepackage{amssymb}
\usepackage{hepnames} %{hepparticles}

\newcommand{\alphas}{\alpha_s}
\newcommand{\RAA}{\rm R_{\rm AA}}
\newcommand{\sqrts}{\sqrt{s}}
\newcommand{\sqrtsnn}{\sqrt{s_{_{\ensuremath{\it{NN}}}}}}
\newcommand{\pp}{p-p}
\newcommand{\pPb}{p-Pb}
\newcommand{\PbPb}{Pb-Pb}

\newcommand{\ttbar}    {\ensuremath{t\bar{t}}}
\newcommand{\bbar}     {\ensuremath{b\bar{b}}}
\newcommand{\jpsi}{\rm J/\Psi}
\newcommand{\qqbar}{q \bar q}
\newcommand{\QQbar}{Q \bar Q}

\newcommand{\Lumi}{\mathcal{L}}
\newcommand{\Lumint}{\mathcal{L}_{\rm int}}

\newcommand{\mH}{m_{_{\rm H}}}
\newcommand{\mtop}{m_{_{\rm top}}}
\newcommand{\mDM}{m_{_{\chi}}}
\newcommand{\mMED}{m_{_{\rm med}}}

\newcommand{\gaga}{\gamma\,\gamma}
\newcommand{\MET}{\not\!\!E_{T}}

\newcommand{\pT}{\rm p_{\rm T}}

\title{CMS physics highlights in the LHC Run 1}

\ShortTitle{CMS physics highlights in the LHC Run 1}

\author{\speaker{David d'Enterria} for the CMS Collaboration\\
CERN, PH Department, 1211 Geneva, Switzerland\\
E-mail: dde@cern.ch}

\abstract{
The main physics results obtained by the CMS experiment during the first three years of operation 
of the CERN Large Hadron Collider (2010--2013, aka. Run 1) are summarized. The advances in our
understanding of the fundamental particles and their interactions are succinctly reviewed
under the following physics topics: (i) Quantum Chromodynamics, (ii) Quark Gluon Plasma, (iii) Electroweak
interaction, (iv) Top quark, (v) Higgs boson, (vi) Flavour, (vii) Supersymmetry, (viii) Dark Matter, 
and (ix) other searches of physics beyond the Standard Model.}

\FullConference{53rd International Winter Meeting on Nuclear Physics,\\
		26-30 January 2015\\
		Bormio, Italy}

\begin{document}

%\scrollmode

\section{Introduction}
%\vspace{0.15cm}

Our theoretical understanding of particle physics is encoded in the Standard Model (SM), a renormalizable
quantum field theory which --unifying quantum mechanics and special relativity-- describes the fundamental
interactions (except gravity) via a local SU$_{\rm c}$(3)$\times$SU$_{_{\rm L}}$(2)$\times$U$_{_{\rm Y}}$(1) 
gauge-symmetry group\footnote{The subindices indicate the conserved {\underline c}olour, and 
{\underline w}eak (hyper)charge, and the action on {\underline l}eft-handed fermions.}. 
The three gauge-symmetry terms give rise to the strong, weak and electromagnetic forces, 
while the particles fall into different representations of these %symmetry  
groups. The SM Lagrangian (without neutrino masses) contains 19 free parameters: 
3 gauge couplings, 9 Higgs--fermion Yukawa couplings,
%6 quark masses, 3 lepton masses,
3 mixing-angles, 2 Charge-Parity (CP) phases, and 2 Higgs-boson couplings, to be determined experimentally.
Despite the fact that the internal consistency and predictive power of the SM have been experimentally
confirmed to great precision in the last 40 years, the theory is not complete and has several outstanding open
questions, 
%that motivated the construction of the CMS~\cite{cms_tdr} (and other experiments) at the LHC~\cite{lhc}, 
%the ultimate particle collider in terms of centre-of-mass (c.m.) energies and luminosity. 
%Among such open problems:
such as:
\begin{enumerate}
\item \underline{Mass generation problem}: The generation of the elementary particles masses through
the Brout-Englert-Higgs (BEH) mechanism~\cite{EBHGHK} (as well as the unitarization of $W\,W$ scattering below $\sim$1~TeV)
required the existence of a new (Higgs) scalar boson which had eluded discovery for over 40 years (until 2012).
\item \underline{Flavour problem}: The huge dominance of matter over antimatter in the Universe cannot
be explained by the known sources of CP violation in the SM. More generally, the SM fails to explain the rationale
behind the observed %semiordered
pattern of fermion families and flavour mixings. %has no apparent explanation.
%Why are there so many types of matter particles and they mix the way they do~?
\item  \underline{Hierarchy / fine-tuning / naturalness problem}: Even with the BEH scalar discovered, the running of its
mass receives power-divergent quantum corrections up to the next known physics scale at Planck energies ($10^{16}$ orders-of-magnitude
above the electroweak scale), unless new particles/symmetries (e.g. Supersymmetry) provide compensating loop corrections.
%Higgs boson compositeness ? new space dimensions ?
\item \underline{Dark matter (DM) problem}: The SM explains only $\sim$4\% of the energy budget of the Universe,
%What weakly-interacting particle accounts for 
the rest being in the form of an unknown DM (plus dark energy), pointing to the existence of %so-far 
new weakly-interacting massive particles (SUSY partners, axions, heavy $\nu$'s,...).
\item \underline{Colour confinement}: Colour-charged particles (quarks and gluons) are always confined inside
SU$_{\rm c}$(3)-invariant hadrons, yet no analytical proof exists that the theory of the strong
interaction (QCD) should be confining. Interesting links of this problem exist with the conjectured (AdS/CFT)
duality between strongly-interacting gauge and string theories~\cite{adscft}.
%The study of QCD matter in heavy-ions collisions provides a testbed
%for a better understanding of colour (de)confinement. %non-perturbative QCD.
%What is the energy evolution of the total hadronic cross sections~? Can one experimentally test the 
%conjectured duality between gauge and string theories (AdS/CFT)~?
%\item \underline{Origin/Nature of the highest-energy cosmic-rays (CRs)}: What are the sources and
%type of particles constituting CRs at energies up to 10$^{20}$ eV ?
\end{enumerate}
Those open problems, among others, motivated the construction of CMS~\cite{cms_det} (as well as other experiments)
at the LHC~\cite{lhc}, the ultimate particle collider in terms of center-of-mass (c.m.) energies ($\sqrts$) and
luminosity ($\Lumi$). This paper succinctly summarizes the progress in our understanding of the SM and the
searches for new physics based on data collected by the CMS experiment during the LHC Run 1 in \pp\
($\Lumint$~=~25~fb$^{-1}$ at $\sqrts$~=~7,~8~TeV), \pPb\ ($\Lumint$~=~34~nb$^{-1}$ at
$\sqrtsnn$~=~5.~TeV), and \PbPb\ ($\Lumint$~=~170~$\mu$b$^{-1}$ at
$\sqrtsnn$~=~2.76~TeV) collisions. %at $\sqrts$~=~7,~8~TeV and $\sqrtsnn$~=~2.76~TeV

%%%%%%%%%%%%%%%%%%%%%%%%%%%%%%%%%%%%%%%%%%%%%%%%%%%%%%%%%%%%%%%%%%%%%%%%%%%%%%%%%%%%%%%%
\section{Quantum Chromodynamics (QCD)}
%\vspace{0.15cm}

\begin{figure}[htpb]
\includegraphics[width=7.2cm,height=6.cm]{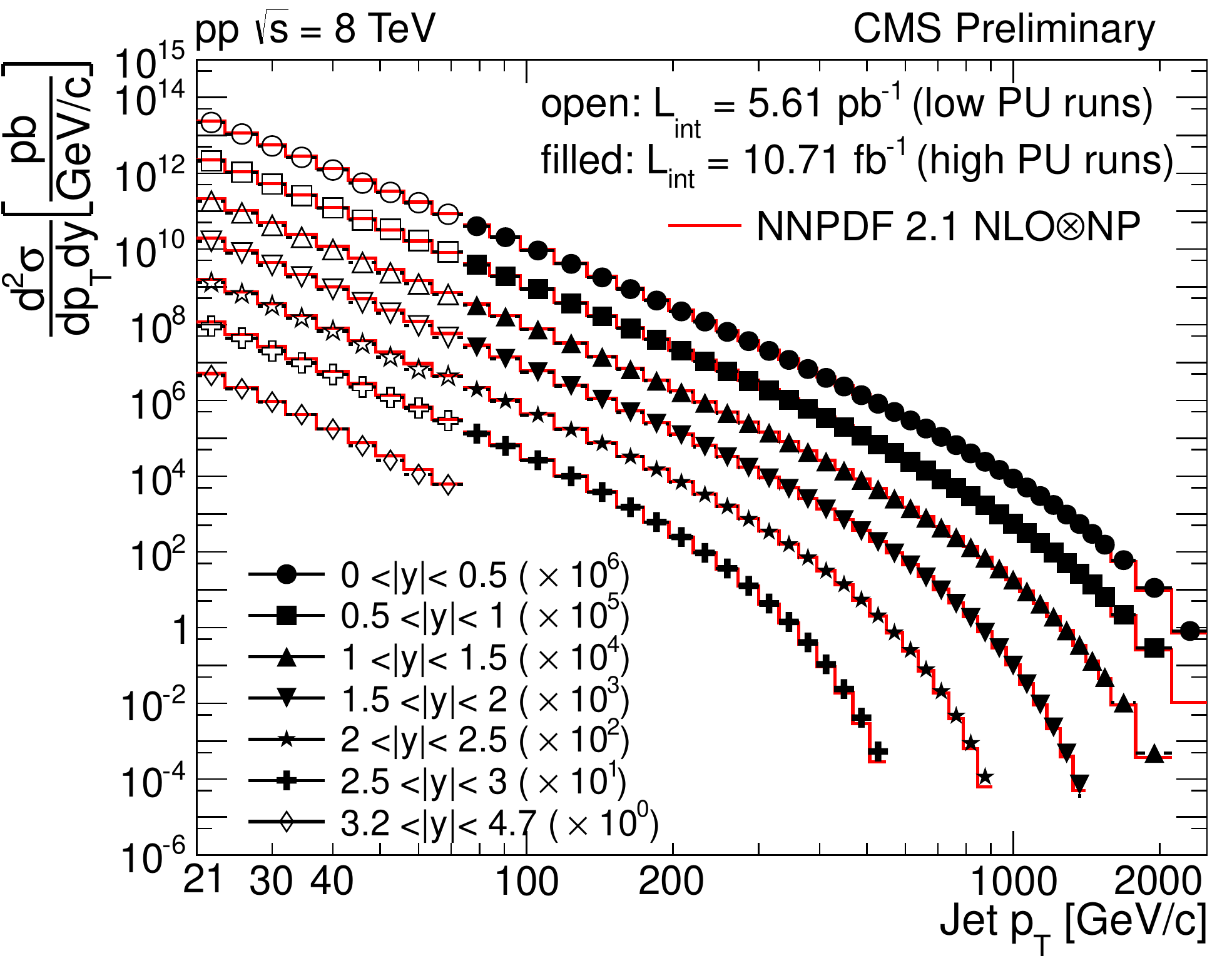}
\includegraphics[width=8.2cm,height=6.cm]{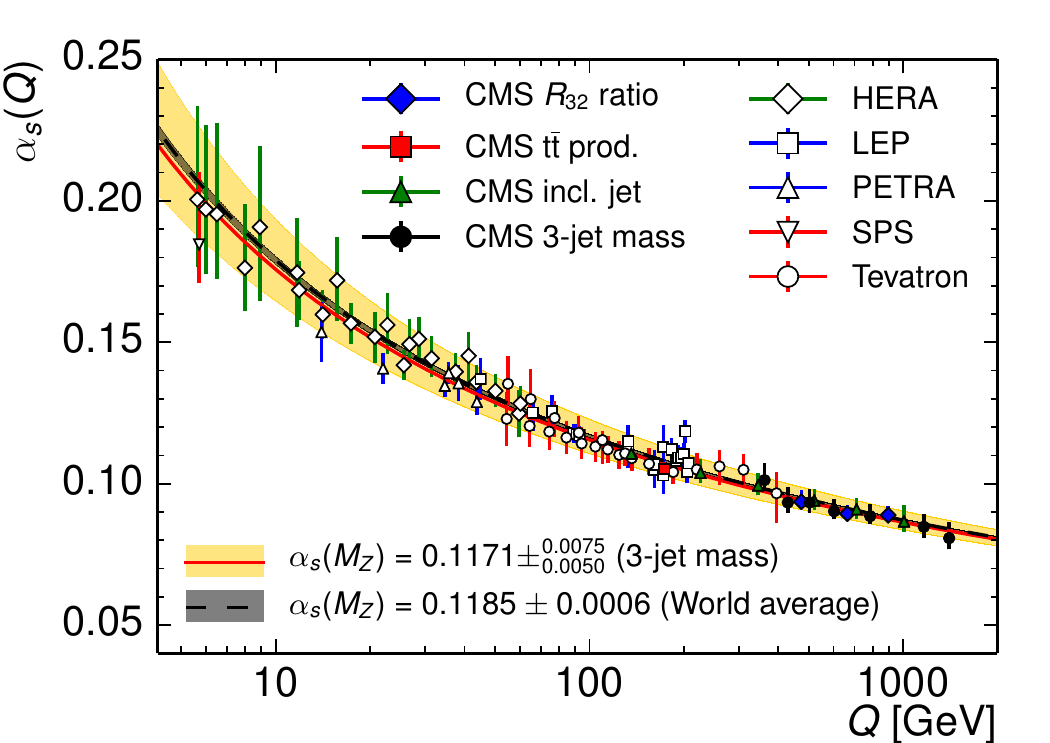}
\caption{Compilation of CMS measurements of $\pT$-differential cross sections for jets in \pp\ at 8~TeV
compared to NLO pQCD predictions~\cite{CMS_FSQ_jets} (left), and strong coupling $\alphas$ versus energy
scale~\cite{CMS:2014mna} (right).}
\label{fig:QCD}
\end{figure}

The processes with the largest cross sections in hadronic collisions are mediated 
by the strong force between the colliding quarks and gluons, described by QCD --a
%Quantum Chromo Dynamics (QCD) 
quantum field theory with a very rich dynamical content
(asymptotic freedom, infrared slavery, approximate chiral symmetry,...).
%, non trivial vacuum topology
%(instantons), strong CP problem, U$_{\rm A}(1)$ axial-vector anomaly,...).
Perturbative QCD (pQCD) calculations are able to accurately describe the production of jets 
(issuing from the hadronization of energetic partons) over an impressive 14 orders-of-magnitude range in their
cross sections~\cite{CMS_FSQ_jets}   
(Fig.~\ref{fig:QCD}, left). In the chiral limit of massless quarks, QCD has a single parameter to be determined
empirically: its coupling $\alphas$, whose current uncertainty (around 0.6\%)~\cite{pdg} makes of it the least precisely
known of all fundamental interaction strengths in nature. Several CMS measurements --such as ratios of 2- to
3-jets, 3-jet masses, inclusive jet and top-quark cross sections~\cite{CMS:2014mna}-- have allowed us to measure $\alphas$ up to
so-far unprobed scales $Q\approx$~2~TeV (Fig.~\ref{fig:QCD}, right).
Our understanding of the unitarization of the pQCD cross sections %via parton saturation and multiple parton interactions, 
in the $\pT\approx$~1--5~GeV %semihard regime
range~\cite{Grebenyuk:2012qp}, dominated by ``minijets'' and multiparton interactions, has also progressed
through different observables~\cite{UE_MPI}. However, no clear deviation from the standard DGLAP 
parton evolution~\cite{dglap}, due to BFKL- and/or parton saturation~\cite{Albrow:2006xt}, has been observed yet. %~\cite{}.
The most inclusive hadronic observable, the inelastic \pp\ cross section --including ``peripheral''
collisions dominated by diffractive interactions which cannot be computed from first-principles QCD-- has been
also measured~\cite{Chatrchyan:2012nj}. This result, plus others on bulk hadron
production~\cite{Khachatryan:2010us,Chatrchyan:2014qka}, %provides valuable constraints on the 
have improved %the models used to determine the nature 
our knowledge of the highest-energy cosmic-rays observed on Earth through extended %particle showers in the atmosphere
air showers~\cite{d'Enterria:2011kw}. 

%%%%%%%%%%%%%%%%%%%%%%%%%%%%%%%%%%%%%%%%%%%%%%%%%%%%%%%%%%%%%%%%%%%%%%%%%%%%%%%%%%%%%%%%
\section{Quark Gluon Plasma (QGP)}
%\vspace{0.15cm}

\begin{figure}[htbp]
%\begin{minipage}{6.0cm}
\includegraphics[width=7.cm]{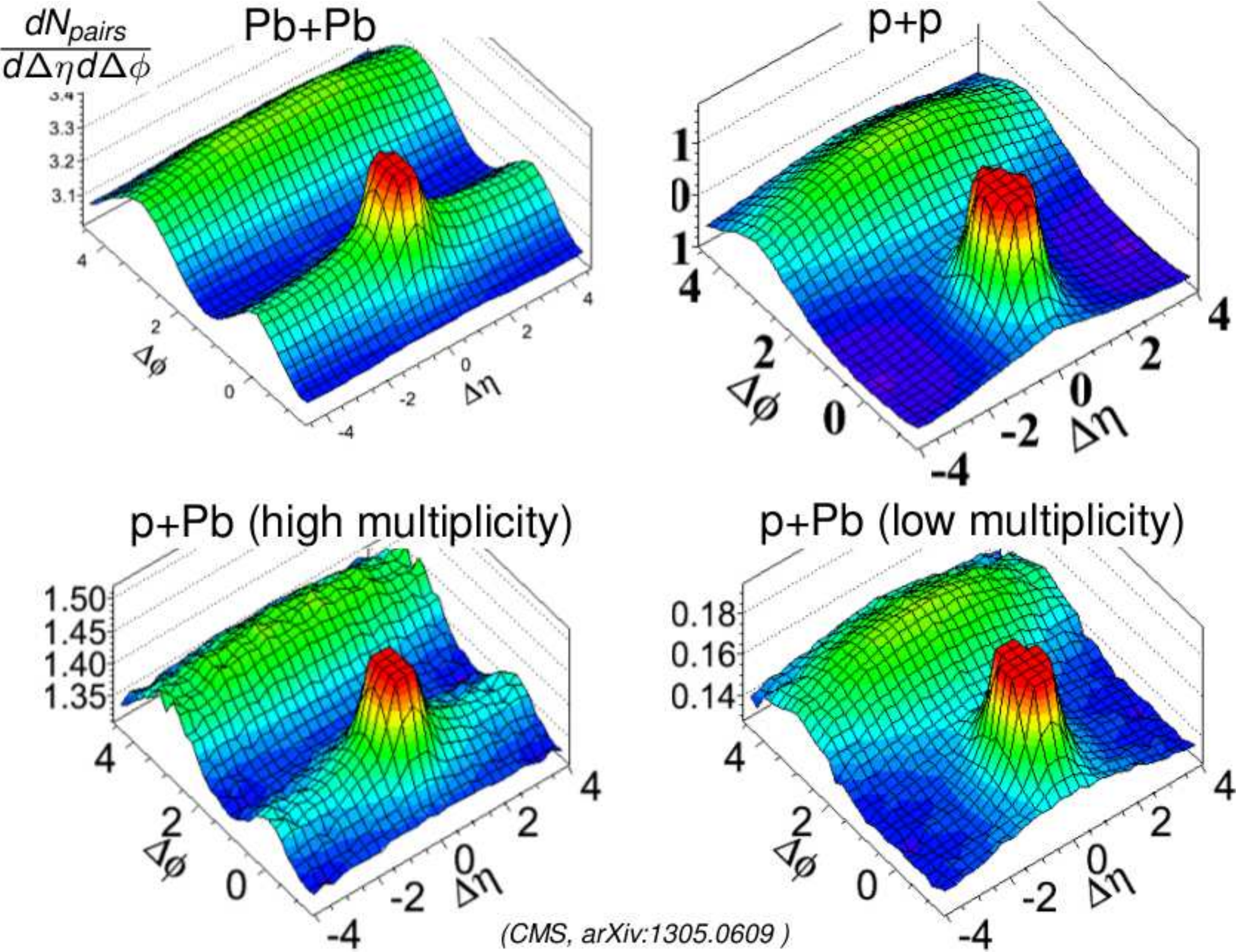}
\includegraphics[width=8.cm,height=5.8cm]{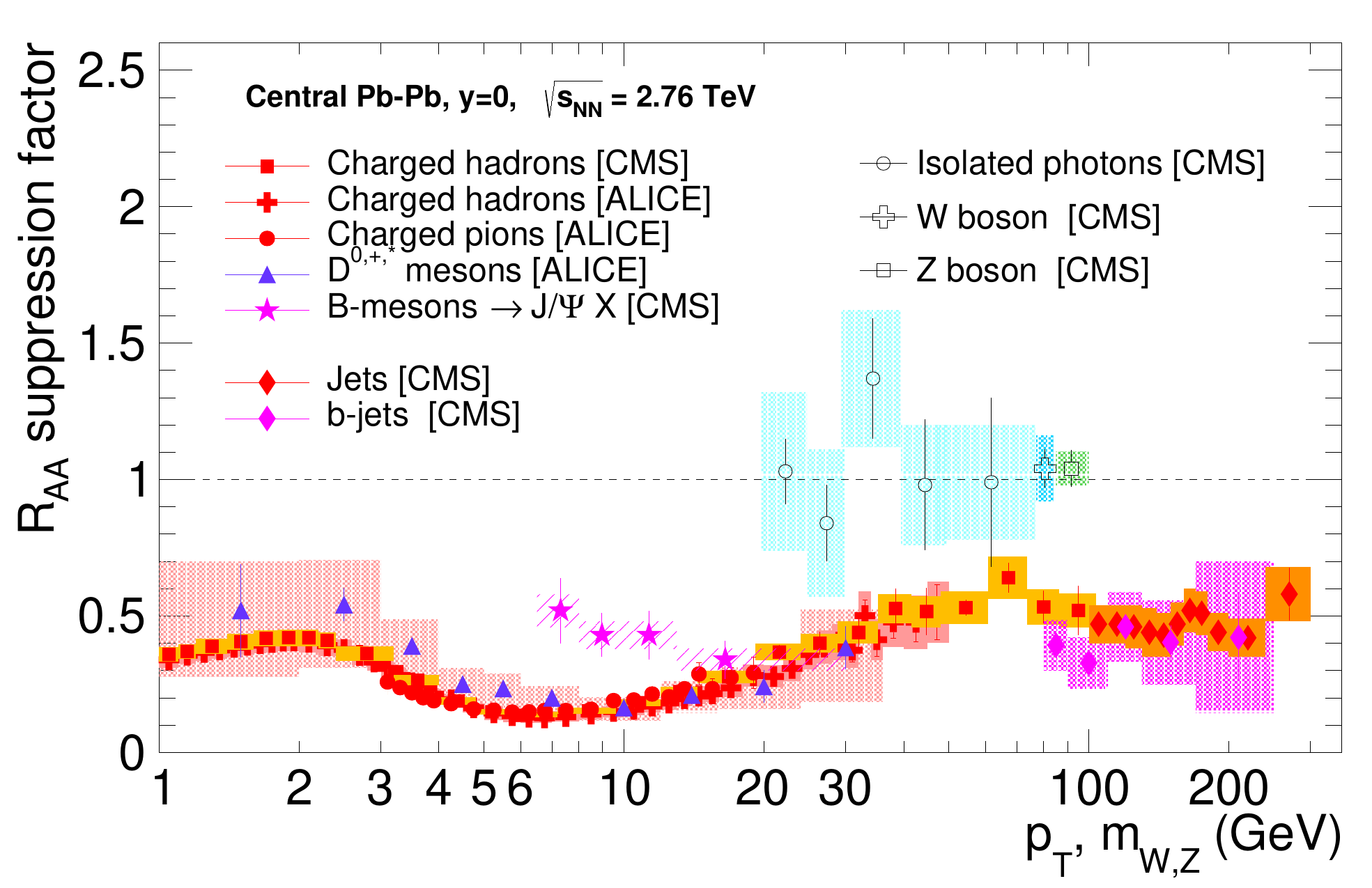}
%\end{minipage}
\caption{Left: Two-particle angular correlation strengths in $\Delta\eta$ vs. $\Delta\phi$ in \pp, \pPb\ and \PbPb\ collisions.
Right: Suppression factors for strongly-interacting particles (hadrons, heavy-quarks, jets) and
weakly-interacting $\gamma, W, Z$ bosons as a function of $\pT$ (or mass, for the weak bosons) in central \PbPb\ collisions.} 
\label{fig:QGP}
\end{figure}

QCD is the only SM sector whose {\it collective} dynamics --phase diagram, (deconfinement and chiral)
phase transitions, thermalization of fundamental fields-- is accessible to scrutiny in the lab %oratory
through the study of the hot and dense partonic medium produced in collisions of heavy nuclei.
Interestingly, the large number of multiparton interactions in ``central'' collisions in the smaller \pp\ and
\pPb\ systems at the LHC produces also final states which share many characteristics of those found in heavy-ion
collisions. One of the surprises of the first LHC run has been the observation of long-range near-side angular
correlations (over $\Delta\eta\approx$~8 units of pseudorapidity) of hadrons 
produced in high-multiplicity \pp, \pPb\ and \PbPb\ collisions (Fig.~\ref{fig:QGP},
left)~\cite{ridge}, whose kinematical properties are consistent with the formation of a dense
parton system (describable with the lattice QCD equation-of-state for a QGP) which expands
hydrodynamically~\cite{Kozlov:2014hya}.

Among all heavy-ion observables, particles with large $p_T$ and/or mass (``hard probes'') are useful
tomographic tools of the produced medium. In the absence of medium effects, their perturbative production
should just be that from an incoherent superposition of independent \pp\ collisions, i.e. the ratio $\RAA$ of
\PbPb\ yields over \pp\ cross sections (normalized by the transverse overlap function of the collision) should
be one. Experimentally, this is the behaviour observed for weakly-interacting probes (such as $\gamma, W, Z$
bosons)~\cite{Chatrchyan:2012vq,Chatrchyan:2012nt,Chatrchyan:2011ua},  
modulo small modifications due to nuclear %parton distribution functions
PDFs, but not so for strongly-interacting
particles (gluons, light and heavy quarks, and their fragmentation
products)~\cite{CMS:2012aa,Chatrchyan:2011sx,Chatrchyan:2013exa} which 
appear suppressed by up to a factor of seven (Fig.~\ref{fig:QGP}, right). The latter is a characteristic
signature of large parton energy loss in the QGP formed in the collision. Intriguing yield suppressions have
been also observed in the $\jpsi$~\cite{Chatrchyan:2012np} and $\Upsilon$~\cite{Chatrchyan:2012lxa} families,
consistent with strong final-state interactions of $\QQbar$ bound states in the plasma.

%%%%%%%%%%%%%%%%%%%%%%%%%%%%%%%%%%%%%%%%%%%%%%%%%%%%%%%%%%%%%%%%%%%%%%%%%%%%%%%%%%%%%%%%
\section{Electroweak physics}
%\vspace{0.15cm}

\begin{figure}[htpb]
\centering
\includegraphics[width=9.cm,height=7.0cm]{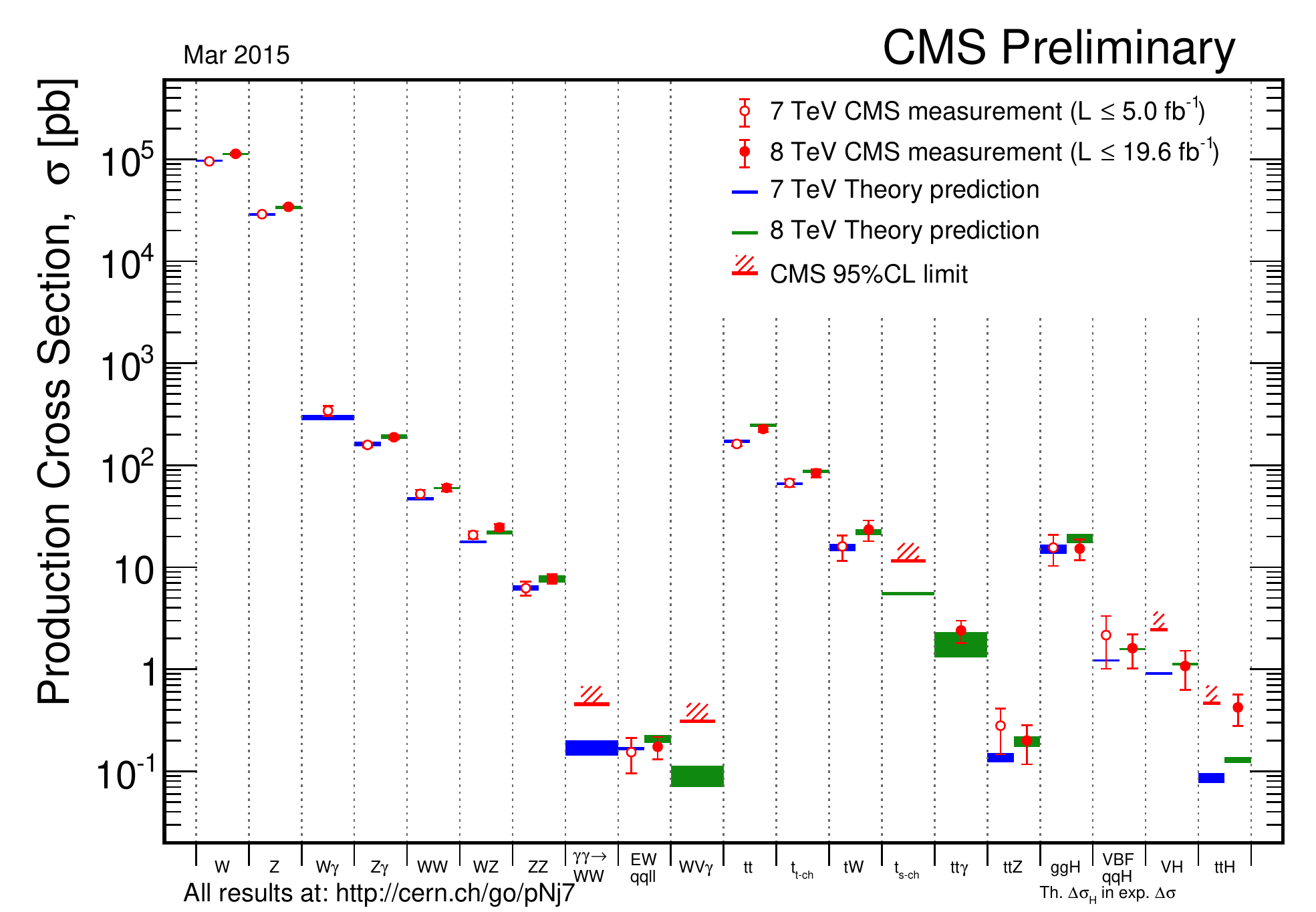}
\includegraphics[width=6.cm,height=6.6cm]{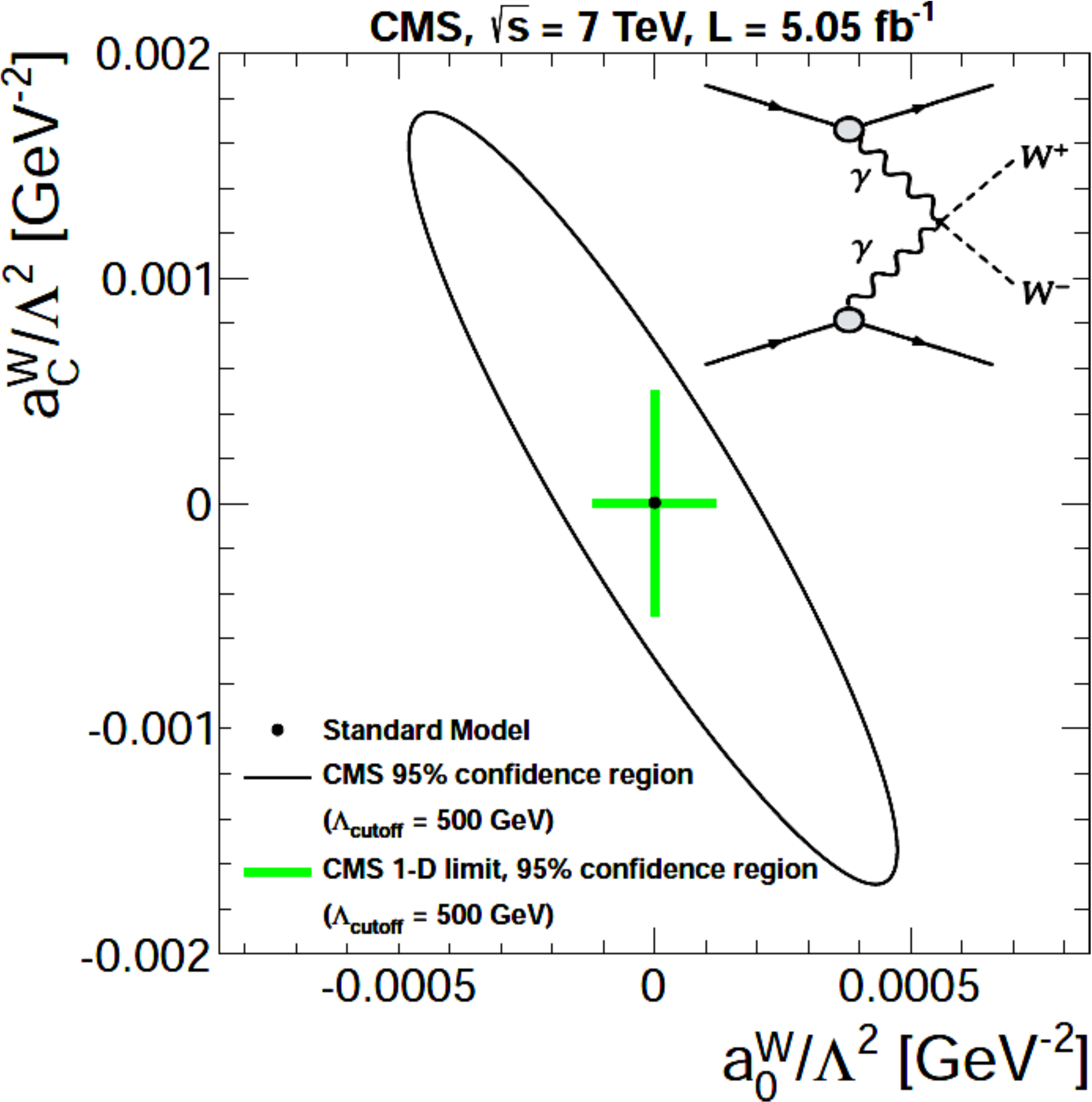}
\caption{Left: Cross sections for various SM processes measured in \pp\ at %collisions at
7,~8~TeV compared to the corresponding (N)NLO theoretical predictions. 
Right: Limits on aQGC from the $\gaga\to W\,W$ measurement~\cite{Chatrchyan:2013foa}.}
\label{fig:EWK}
\end{figure}

The electroweak sector of the SM describes processes involving the $\gamma, W, Z$ (and Higgs, see later)
bosons. Thanks to their precisely-known theoretical production cross sections, the differential distributions
of $W$ and $Z$ bosons (aka. ``standard candles'') have improved our knowledge of the flavour dependence of
the quark densities in the proton~\cite{jrojo}. In addition, many other electroweak cross sections have been
measured with good precision, down to the hundreds-of-fb scale, finding  excellent agreement with
next-to-leading (NLO) or next-to-NLO theoretical predictions %(left half of 
(Fig.~\ref{fig:EWK}, left). Multiple first-ever measurements are worth to highlight: $W+t$~\cite{Chatrchyan:2012zca}, 
$\ttbar$+$\gamma$~\cite{CMS:2014wma}, $\ttbar+Z$~\cite{Khachatryan:2014ewa}, $\gaga\to W\,W$~\cite{Chatrchyan:2013foa}, 
and vector-boson-fusion (VBF) $Z$ boson production~\cite{Khachatryan:2014dea}. Particularly important are
the processes where two or more bosons are produced, which provide also novel stringent limits on anomalous
triple~\cite{aTGQ} and quartic~\cite{Chatrchyan:2013foa,aQGQ} gauge couplings (aQGC, Fig.~\ref{fig:EWK}, right).

%%%%%%%%%%%%%%%%%%%%%%%%%%%%%%%%%%%%%%%%%%%%%%%%%%%%%%%%%%%%%%%%%%%%%%%%%%%%%%%%%%%%%%%%
\section{Top quark physics}
%\vspace{0.15cm}

The top quark, being the heaviest elementary particle, features the strongest coupling to the Higgs field.
Its mass is thus a fundamental SM parameter with far-reaching implications on naturalness, stability of the
electroweak vacuum, SUSY predictions for $\mH$, etc. The large  top-pair production cross sections at the 
LHC provide large data samples to study its properties 
\begin{figure}[htbp!]
\includegraphics[width=8.1cm,height=7.cm]{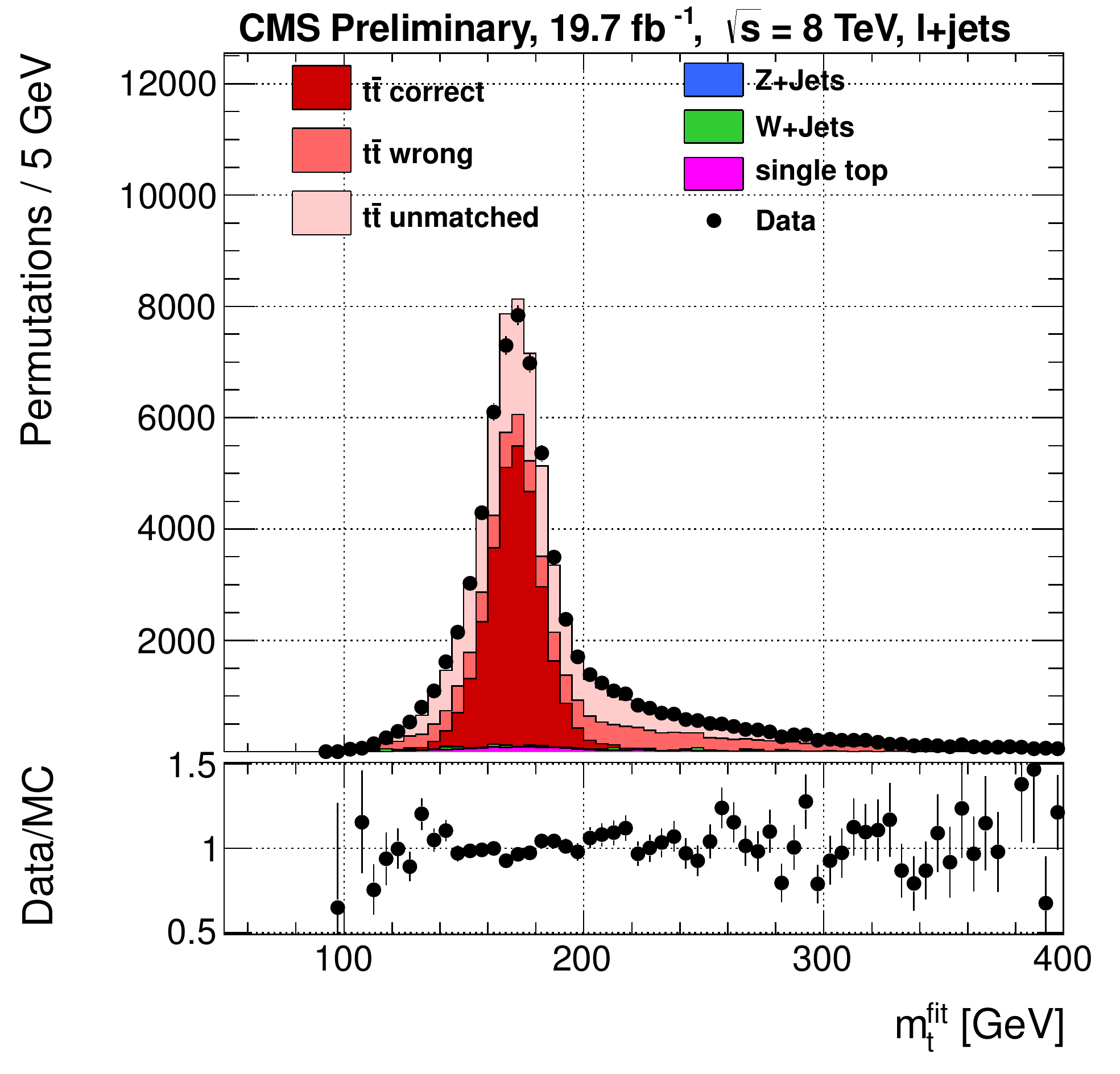}
\includegraphics[width=6.9cm,height=7.cm]{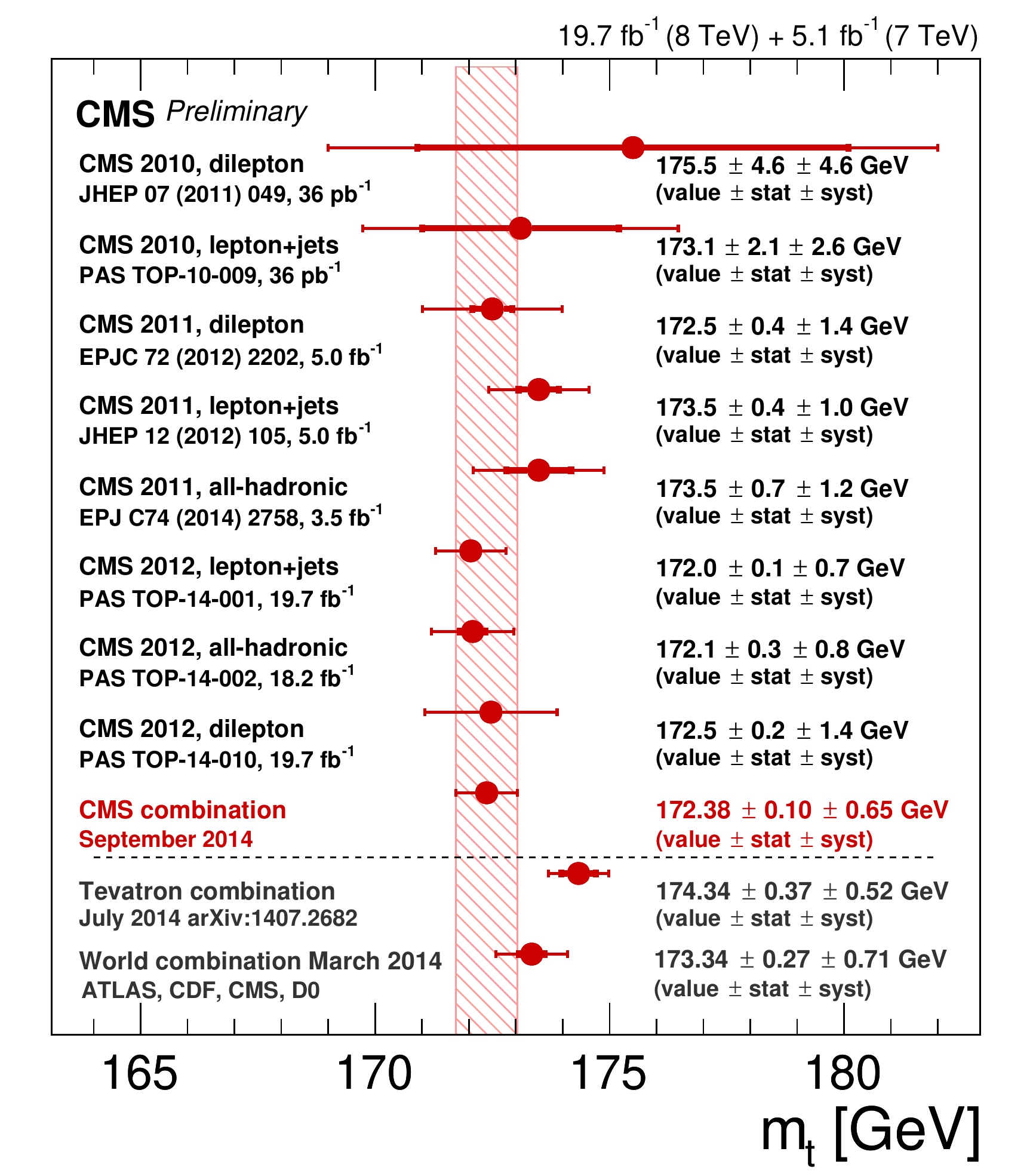}
\caption{Left: $\ttbar\to(W\,W\,b\,b)\to\Plepton$+jets invariant mass distribution. 
Right: Combined $\mtop$ measurements.}
\label{fig:top}
\end{figure}
in various $\ttbar\to W\,W\,b\,b$ final states depending on the decays of the two $W$ bosons (fully-hadronic,
lepton+jets, or fully-leptonic). The most precise $\mtop$ single measurement to date is that obtained from a
kinematic fit of $\ttbar$ events decaying into a lepton plus %electron or muon and 
at least four jets (Fig.~\ref{fig:top}, left)~\cite{cmstop2014-001}. The simultaneous
data fit with an overall jet energy scale factor (JSF), constrained by the known mass of the $W$ boson in
$\qqbar$ decays, yields $\mtop$~=~172.04$\pm$0.19~(stat+JSF)~$\pm$0.75~(syst)~GeV. 
The combined CMS measurements yield a top mass with a 0.4\% uncertainty, $\mtop$~=~172.38~$\pm$~0.67~GeV
(Fig.~\ref{fig:top}, right). Such a value is about 2$\sigma$ higher than the $\mtop\lesssim$~171~GeV
required for the stability of the electroweak 
vacuum given by the evolution of the Higgs quartic coupling (in the absence of new physics up to the Planck
scale) for $\mH=$~125~GeV, and $\alphas=$~0.1184~\cite{Buttazzo:2013uya}, indicating that the current
Universe is in a metastable state. It is interesting to note that this result hinges partly on our limited
understanding of a non-pQCD phenomenon --the modelling of the colour reconnection among the decay
partons of the top-quarks and the partons from the rest of the \pp\ event-- which constitutes one of the
leading theoretical uncertainties on $\mtop$. Beyond $\ttbar$, %many measurements have been also carried out of
single-top cross sections have been measured in the $W+t$ associated production~\cite{Chatrchyan:2012zca} and
in the $t$-channel~\cite{Khachatryan:2014iya} allowing the extraction (via
$\sigma_{\rm single-t}\propto\left|V_{\rm tb}\right|^2$) of the Cabibbo-Kobayashi-Maskawa (CKM)  $t$-$b$ element,
$|V_{\rm tb}|=$~0.998~$\pm$~0.038~$\pm$~0.016, independently of assumptions on the number of quark generations
and unitarity of the CKM matrix.

%%%%%%%%%%%%%%%%%%%%%%%%%%%%%%%%%%%%%%%%%%%%%%%%%%%%%%%%%%%%%%%%%%%%%%%%%%%%%%%%%%%%%%%%
%
\section{Higgs boson physics}
%\vspace{0.15cm}

The main driving force for the construction of the LHC was to close the last missing piece of the
SM: the generation the $W,Z$ bosons (as well as fermions') masses through the spontaneous breaking of the
SU$_{\rm L}$(2)$\times$U$_{\rm Y}$(1) symmetry of the Lagrangian by the presence of a new (Higgs) scalar
doublet. 
\begin{figure}[htbp]
\includegraphics[width=7.5cm,height=7.5cm]{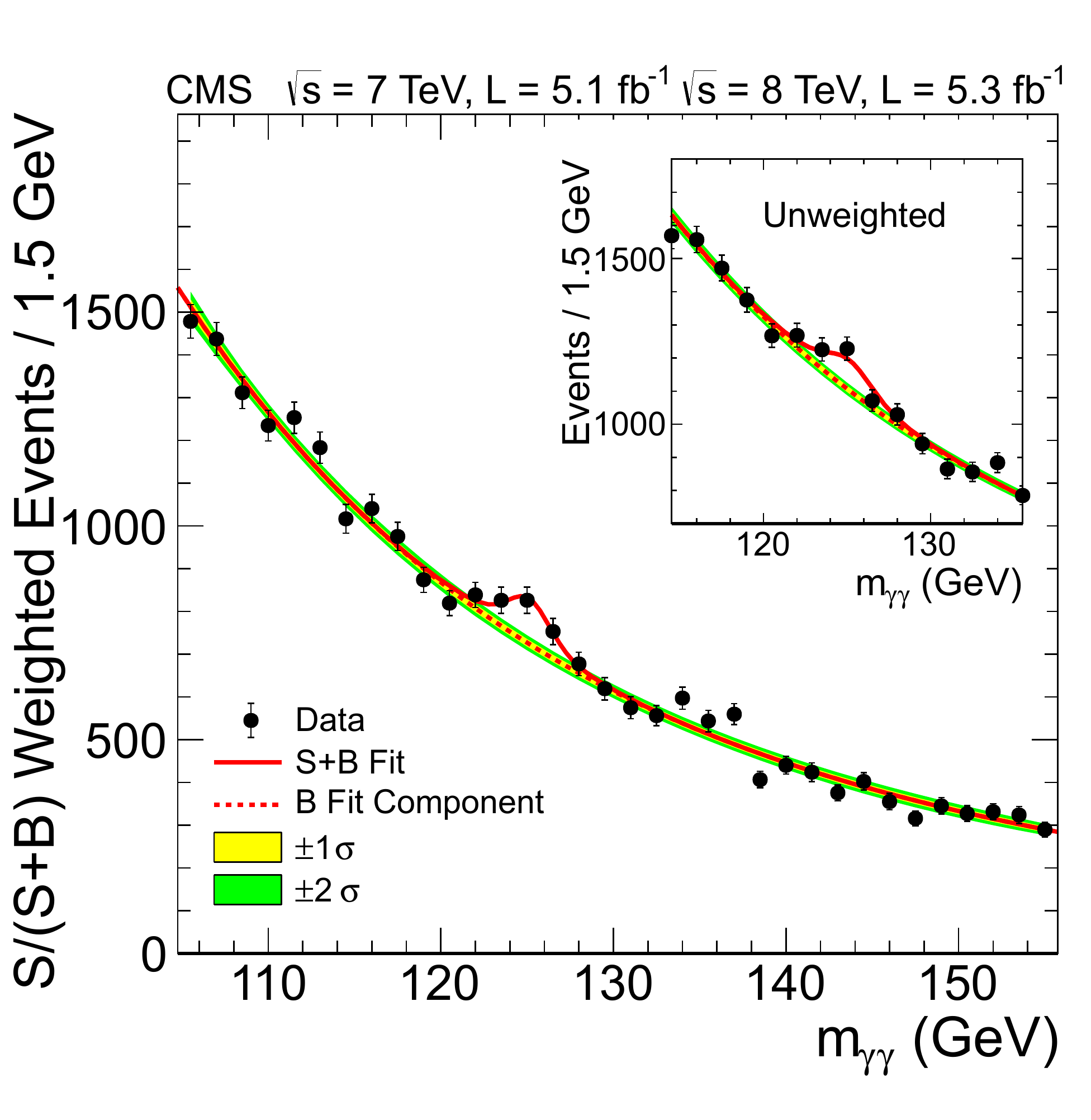}
\includegraphics[width=7.8cm]{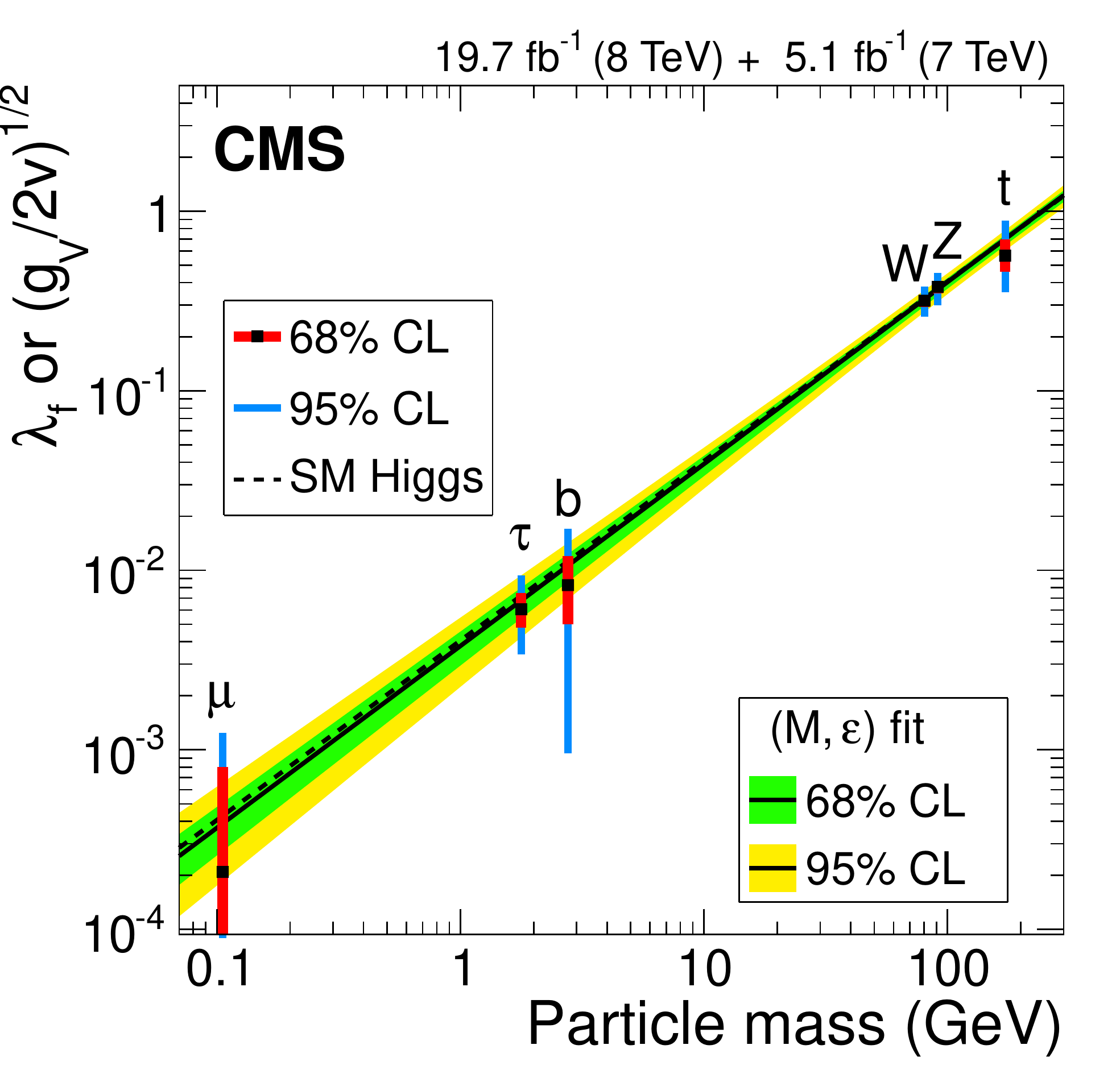}
\caption{Left: Diphoton invariant mass distribution with the Higgs signal over the continuum background~\cite{Khachatryan:2014ira}.
Right: Higgs boson couplings to fermions and bosons as a function of the particle mass~\cite{Khachatryan:2014jba}.}
\label{fig:higgs}
\end{figure}
Higgs boson searches at the LHC involve a large combination of production channels: $g$-$g$ fusion, VBF
fusion, and associated with $W,Z$,top; and decay modes: $\gaga$ and $Z\,Z^\star\to$4$\Plepton$ providing
the ``cleanest'' signal, and $W\,W^\star,\bbar,\tau\,\tau$ the largest cross sections. The scalar boson was
first observed in the H$\to\gaga,Z\,Z^\star(4\Plepton)$ modes~\cite{Chatrchyan:2012ufa} as a resonance over 
smooth continuum backgrounds (Fig.~\ref{fig:higgs}, left), and then as broader (2--4)$\sigma$ excesses in the
$W\,W^\star$~\cite{Chatrchyan:2013iaa} and 3rd-family fermion  
($\bbar,\tau\,\tau$)~\cite{Chatrchyan:2014nva,Chatrchyan:2014vua} decays. The combination of many production
and decay channels confirm that the new resonance couples proportionally to the (Yukawa)
fermions' masses as well as to the square of the weak-boson masses (Fig.~\ref{fig:higgs},
right)~\cite{Khachatryan:2014jba}. Its mass, $\mH$~=~125.09~$\pm$~0.21~$\pm$~0.11~GeV, obtained combining the
high-resolution CMS+ATLAS channels, is known already with an impressive uncertainty below 0.2\%~\cite{Aad:2015zhl}. 
Its width has been constrained through the ratio of the on- and off-shell $ZZ^\star(4\Plepton)$ decays,
and found to be smaller than 5.4 times the SM prediction: $\Gamma_{_{\rm H}}<$~22~MeV (95\% CL)~\cite{Khachatryan:2014iha}.
Its quantum numbers ($J^{PC}=0^{++}$), determined mostly through the kinematical distributions of the
$ZZ^\star$ decay leptons, are those expected for a SM Higgs boson~\cite{Khachatryan:2014kca}. 
Taken together, all measured properties show no deviation so-far from the expectations for the SM
Higgs boson.

%%%%%%%%%%%%%%%%%%%%%%%%%%%%%%%%%%%%%%%%%%%%%%%%%%%%%%%%%%%%%%%%%%%%%%%%%%%%%%%%%%%%%%%%
\section{Flavour physics}
%\vspace{0.15cm}

\begin{figure}[htbp]
\centering
\includegraphics[width=4.cm,height=5.2cm]{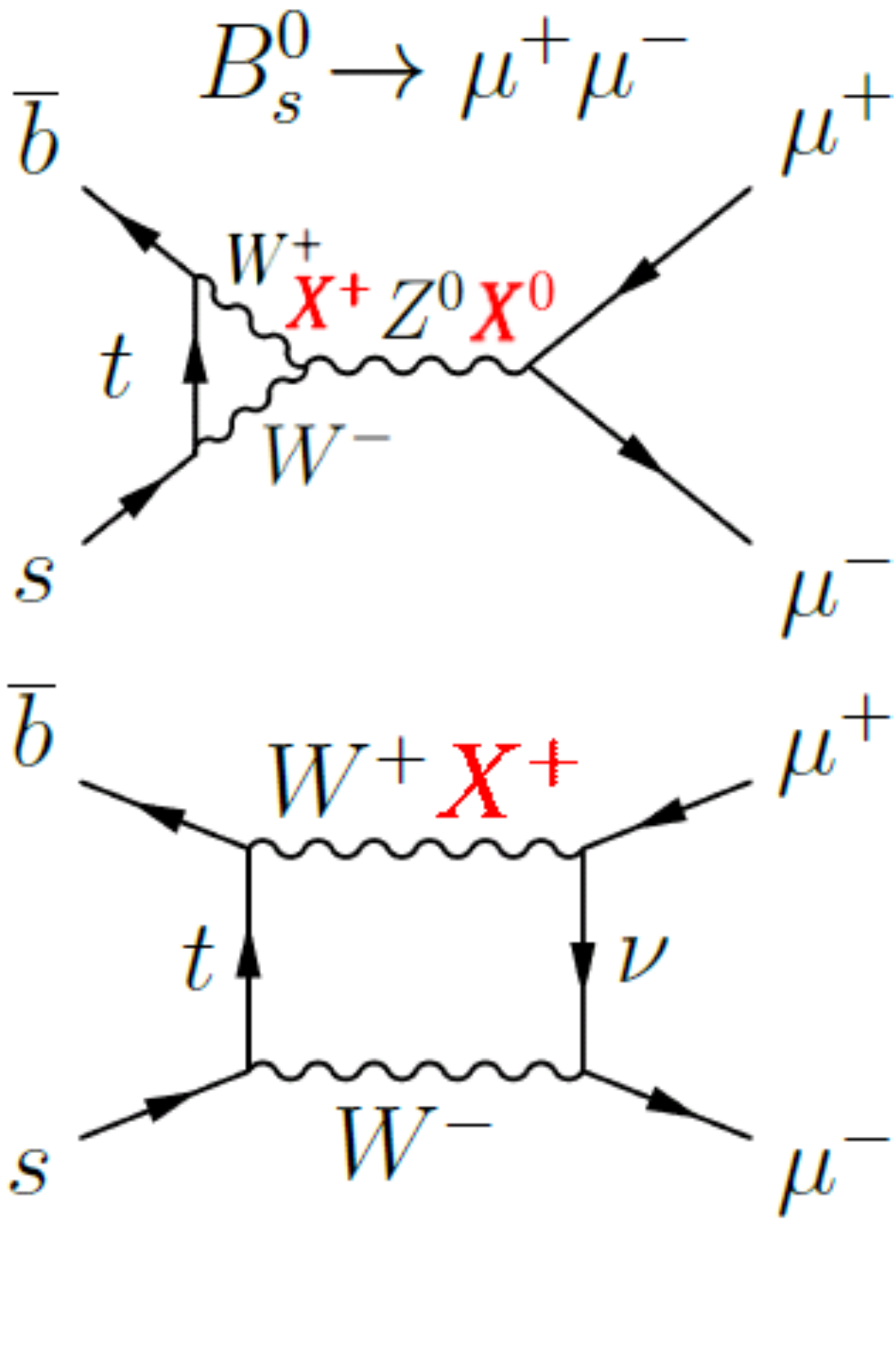}
\hspace{0.8cm}
\includegraphics[width=9.8cm,height=5.65cm]{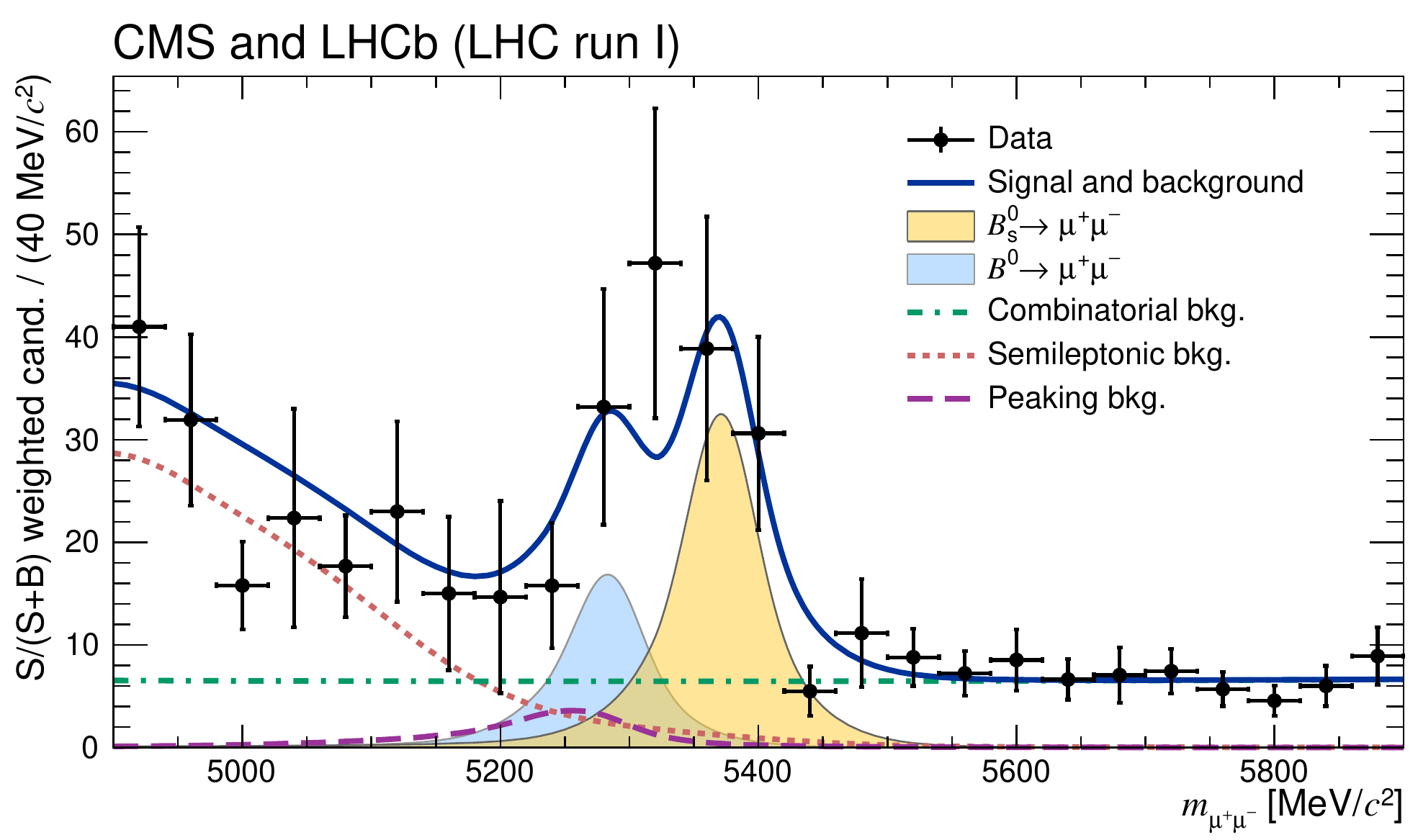}
\caption{Left: Higher-order FCNC diagrams for the SM $B^0_s$ decay 
(and in SM extensions with new particles $X^{0,\pm}$ altering the decay rate).
Right: Measured invariant dimuon mass in the $B^0_s,B^0$ decay range~\cite{CMS:2014xfa}.
\label{fig:Bsmumu}}
\end{figure}

The known differences between particles and antiparticles in the SM, induced by the CP-violation of the
electroweak interaction, are way too small to explain the observed matter-antimatter imbalance
and new particles and/or CP-phases are needed in order to explain how baryon dominance, 
$(n_{_{\rm B}}-n_{_{\rm \bar{B}}})/n_{_{\gamma}}\approx 10^{-9}$, appeared in the
Universe (baryogenesis). Precision flavour studies at the LHC involve
indirect searches of new virtual particles contributing to higher-order loops (Penguin or box, see
Fig.~\ref{fig:Bsmumu}, left), in particular in flavour-changing neutral current (FCNC) processes %and in rare decays
involving $B$-mesons (e.g. $b\to s$ transitions) which are less constrained by lower-energy experiments.
The very-rare $B^0_s,B^0\to\mu^+\mu^-$ decays, with branching ratios BR~$\approx 10^{-8.5,-10}$, have been for years
considered as ``golden channels'' to look for deviations from the SM due to new virtual
contributions. The combined measurement of the CMS and LHCb experiments (Fig.~\ref{fig:Bsmumu}, right)
has established conclusively the existence of $B^0_s\to\mu^+\mu^-$ with a 6.2$\sigma$ statistical
significance (as well as a 3$\sigma$-evidence for the $B^0$ decay) with a BR fully
compatible with the SM prediction~\cite{CMS:2014xfa}. 
Such a result imposes novel constraints on the flavour-changing sector of any viable theoretical extension of the SM.
\section{Supersymmetry (SUSY)}
%\vspace{0.15cm}

SUSY has been for decades a leading candidate theory for the extension of the SM due to various reasons: 
% The existence of new SUSY partners, differing by 1/2 unit of spin, for every SM particle
%it provides a simple way to stabilize the running of the Higgs mass, 
%as their amplitude in the quantum corrections come with opposite sign and cancel the SM contributions 
(i) it solves the hierarchy problem by providing new spartners, differing by 1/2 spin-unit for every SM particle,
whose quantum corrections %compensate the SM contributions to 
stabilize the running of the Higgs mass, (ii) it provides viable dark matter candidates (in
R-parity-conserving SUSY) in the form of stable lightest SUSY Particles (LSP, such as neutralinos or
gravitinos), (iii) it leads to the high-energy unification of the three interaction couplings, plus 
(iv) it has various theoretically-appealing features (simplest extension of the Poincar\'e space-time
symmetry, fermion-boson symmetry required by string theory,...). 
Most of the SUSY searches are based on the assumption of R-parity conservation (leading to sparticle 
pair-production which decay into other sparticles plus any number of SM particles) with final states
characterized by large missing transverse energy ($\MET$) from the invisible LSP ($\tilde{\chi}^0$) %_1$)
at the end of a decay chain, plus multi-jets,$\gamma$ and/or same-sign leptons from intermediate sparticle
decays. 
\begin{figure}[htbp]
\includegraphics[width=15.0cm]{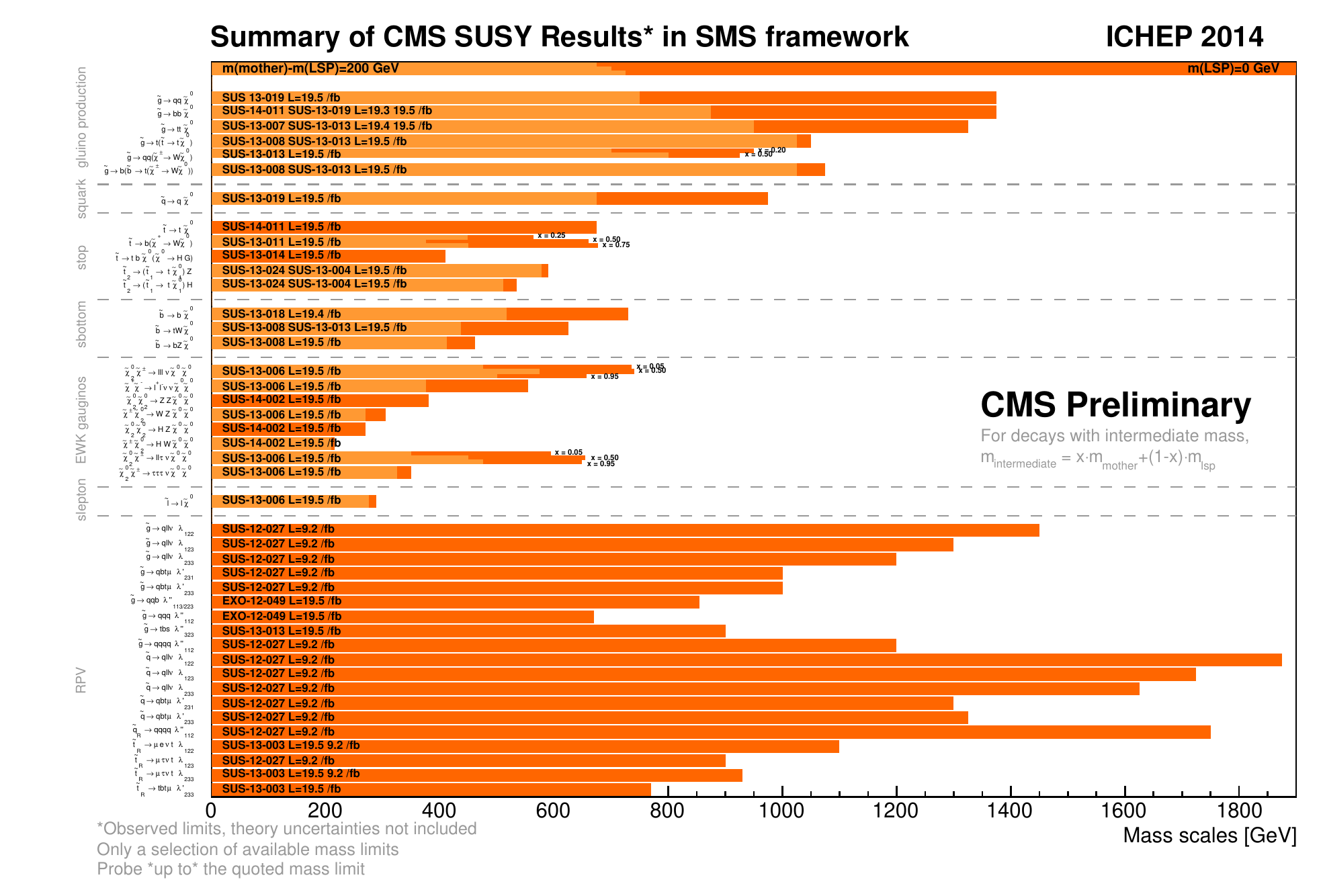}
\caption{Summary of exclusion limits on (SMS) SUSY particles masses in different CMS searches.}
\label{fig:susy}
\end{figure}
Any final excess (or lack thereof) is then interpreted phenomenologically in terms of simplified model space
(SMS) SUSY realizations with a few free parameters (e.g. within constrained-MSSM or mSUGRA with $\tan\beta$,
A, sign($\mu$) plus common scalar $m_0$ and fermion masses $m_{1/2}$ defined at the GUT scale and evolved 
down in energy). Figure~\ref{fig:susy} summarizes the current exclusion limits on spartners masses:
strongly-interacting gluinos and squarks are pushed above $\approx$1~TeV, stops and sbottoms above 
$\approx$0.5~TeV, and electroweak gauginos and sleptons above $\approx$0.3~TeV. 
Such masses, increasingly away from the electroweak scale, render SUSY less and less ``natural''
(i.e. relevant for the resolution of the SM fine-tuning problem).

%%%%%%%%%%%%%%%%%%%%%%%%%%%%%%%%%%%%%%%%%%%%%%%%%%%%%%%%%%%%%%%%%%%%%%%%%%%%%%%%%%%%%%%%
\section{Dark matter (DM)}
%\vspace{0.15cm}

The existence of dark matter, accounting for about 27\% of the Universe energy budget~\cite{Ade:2013sjv}, 
has been confirmed by many observations: (i) non-Keplerian star (galaxies) orbits in
galaxies (clusters), (ii) offset in the distribution of matter observed via gravitational-lensing and via
radiation in colliding clusters of galaxies, (iii) pattern of the power spectrum of the temperature
fluctuations of the cosmic microwave background, and (iv) simulations of the large-scale structure of the
cosmos, among others. All we know of DM is that is sensitive to gravitation, stable, and an early-Universe relic. The
preferred candidate is a Weakly-Interacting Massive Particle (WIMP) of mass $\mDM\approx$~10--1000~GeV with
electroweak-like DM-SM interaction strength, so that its early-Universe annihilation cross sections
($\sigma_{\rm anni}\propto g^4_{\rm ewk}/\mDM$) are compatible with the current DM density.
Many SM extensions include DM candidates such as the LSP in SUSY,
%lightest SUSY Particle (LSP, e.g. neutralinos or gravitinos), 
the lightest Kaluza-Klein tower in extra dimension models, heavy R-handed (sterile) neutrinos,
axions, or particles from a new hidden sector (e.g. moduli fields from string/M-theory compactifications).
The null $\MET$ excesses observed in the Run-1 searches (see previous Section) seem to exclude the
simplest LSP candidates from R-parity-conserving SUSY. %Relaxing the SUSY searches constraints, 
\begin{figure}[htbp]
\includegraphics[width=7.5cm,height=6.5cm]{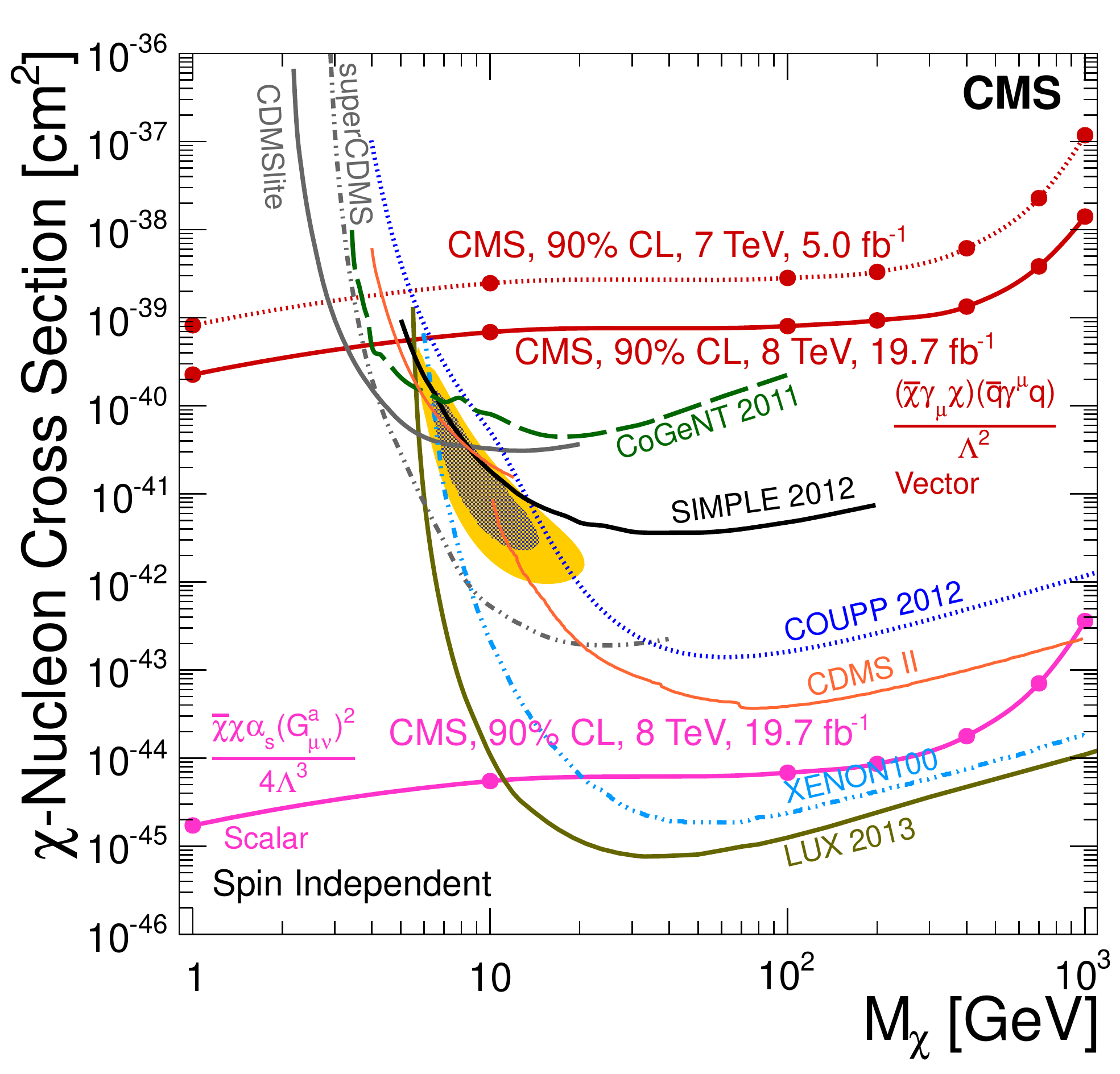}
\includegraphics[width=7.5cm,height=6.5cm]{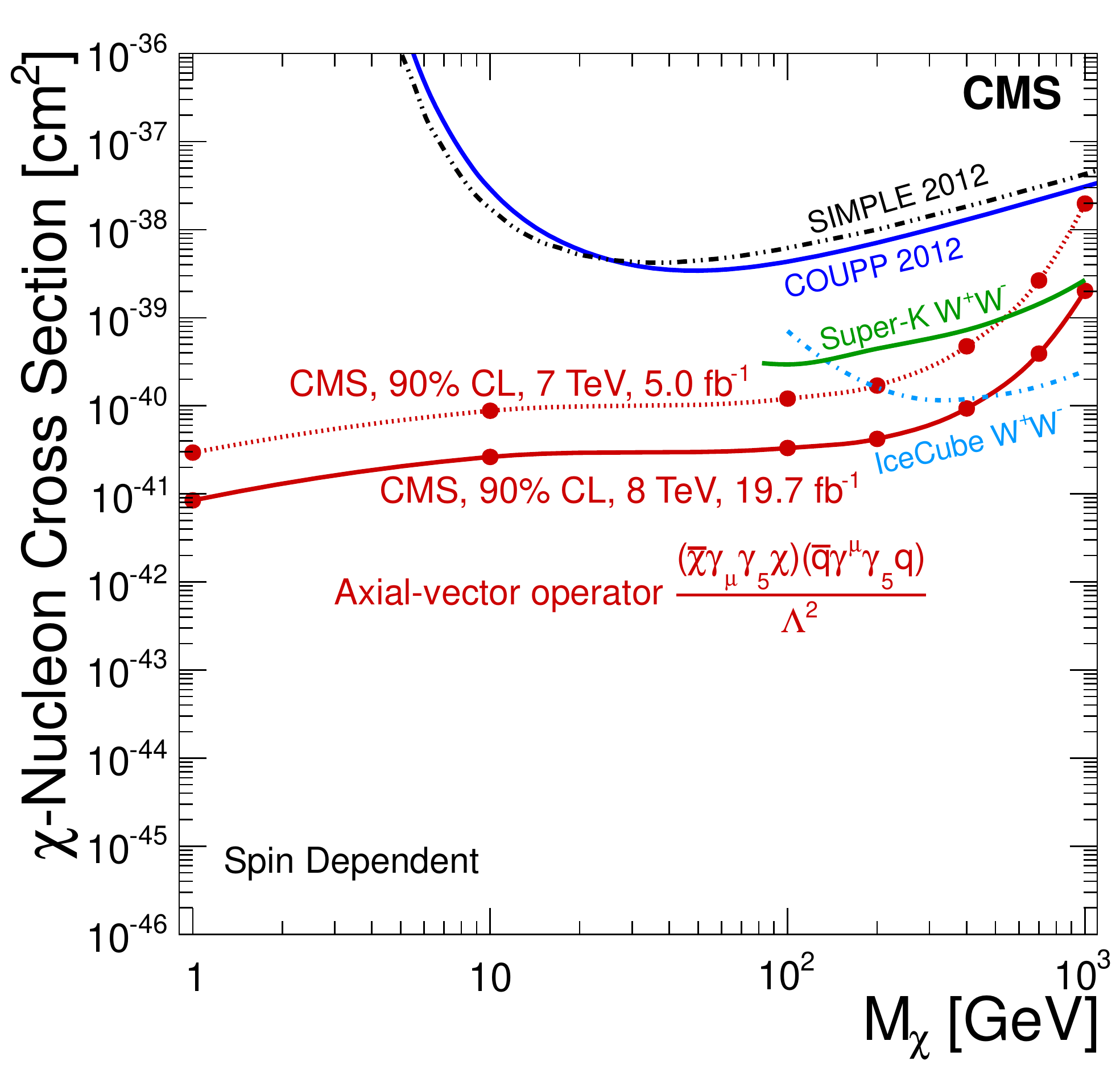}
\caption{Upper limits (90\% CL) on the DM-nucleon cross section vs. DM mass (CMS and direct DM searches) 
for vector and scalar (left) and axial-vector (right) operators~\cite{Khachatryan:2014rra}.}
\label{fig:DM}
\end{figure}
The most generic DM search at colliders involves an unbalanced mono-$X$ final-state where the colliding quarks
or gluons emit an object $X$=jet,photon,$W$,... prior to their annihilation into a pair of
DM particles (indirectly observed via $\MET$). A third DM search approach at the LHC (for $\mDM<\mH/2$)
involves looking for invisible Higgs decays~\cite{Chatrchyan:2014tja}.
The lack (so far) of mono-jet,photon excesses above the SM backgrounds (most notably
$Z(\nu\nu$)+jet,$\gamma$) can be interpreted within an effective field theory (EFT) for the SM-DM interaction
--characterized by an effective scale $\Lambda=\mMED\sqrt{g_{_{\chi}}\,g_{_{q,q}}}$ (where $\mMED$ is the 
mediator mass and $g_{_{\chi,q,g}}$ its couplings to DM and SM particles), and $\mDM$. Corresponding limits
in the plane of DM-nucleon cross section $\sigma(\chi\,N)$ %\to\chi\,N)\propto\nu_{\chi\,N}^2/\Lambda^2$, where 
%$\nu_{\chi\,N}$ is the DM-nucleon reduced mass) 
vs. $\mDM$ can be set and compared to direct underground searches (Fig.~\ref{fig:DM}). The current LHC limits,
$\sigma(\chi$-N)~$\lesssim 10^{-40}$~cm$^{2}$ (for spin-independent interactions, left) and $10^{-41}$~cm$^{2}$
(for spin-dependent ones, right) are particularly competitive at low DM masses ($\mDM\lesssim$~10~GeV) where
the tiny recoil energy of nuclei in direct underground searches is not visible, but the collider $\MET$
searches benefit from potentially large Lorentz boosts.

%%%%%%%%%%%%%%%%%%%%%%%%%%%%%%%%%%%%%%%%%%%%%%%%%%%%%%%%%%%%%%%%%%%%%%%%%%%%%%%%%%%%%%%%
\section{Other searches of physics beyond the Standard Model (BSM)}
%\vspace{0.15cm}

Apart from the aforementioned dedicated SUSY and dark matter studies, almost a hundred other BSM searches have
been carried out in \pp\ collisions at 7,8~TeV looking for excesses over the SM predictions, mostly in the %single high-$\pT$ and/or
high invariant-mass tails of the distributions of pairs of objects (jets, leptons, photons,...).
\begin{figure}[htbp]
\includegraphics[width=15.0cm]{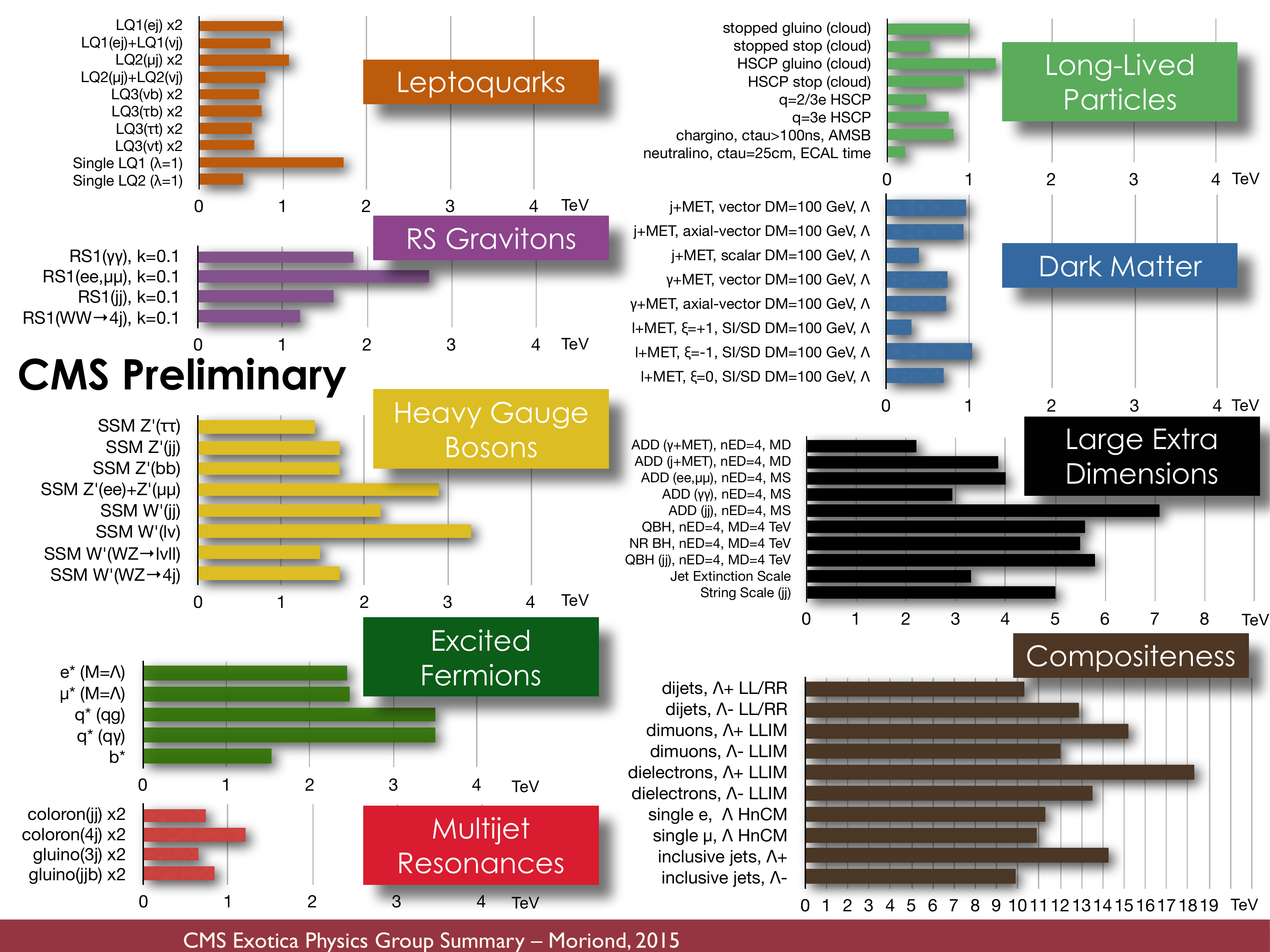}
\caption{Summary of CMS limits on new physics particle-masses/scales in different BSM searches.}
\label{fig:bsm}
\end{figure}
Figure~\ref{fig:bsm} summarizes the latest (95\% CL) limits on the mass of new particles (m$_X$) 
or the scale of new physics ($\Lambda$). The highest scales probed ($\Lambda\gtrsim$~15~TeV) correspond to
searches of quark compositeness (contact interactions)~\cite{Chatrchyan:2013muj}, followed by $\Lambda\gtrsim$~5--7~TeV
for virtual graviton exchanges~\cite{Khachatryan:2014cja} based on the Arkani-Hamed--Dimopoulos--Dvali (ADD) model
of large extra spatial dimensions~\cite{ADD}. Heavy gauge bosons ($Z', W'$)~\cite{Khachatryan:2014tva} --from
new U(1), SU(2) gauge symmetries at high energies-- or new excited fermions ($\Plepton^\star$,
$q^\star$)~\cite{Khachatryan:2014aka} are excluded below masses m$_X\approx$~1.5--3.5~TeV 
(depending on their decay channels). %The mass of gravitons from 
The scale for the onset of quantum gravity, in Randall-Sundrum (RS) warped extra-dimension
scenarios~\cite{RS}, is pushed above the $\Lambda\approx$~2.5~TeV range~\cite{Chatrchyan:2012kc}, whereas 
leptoquarks~\cite{Khachatryan:2015bsa}, long-lived particles~\cite{Khachatryan:2015jha} (e.g. from
R-parity-violating SUSY), and fourth-generation b',t' quark partners~\cite{Chatrchyan:2013wfa} are excluded
for masses below m$_X\approx$~0.6~TeV.

%\clearpage
%%%%%%%%%%%%%%%%%%%%%%%%%%%%%%%%%%%%%%%%%%%%%%%%%%%%%%%%%%%%%%%%%%%%%%%%%%%%%%%%%%%%%%%%
\section*{Summary}

The main physics results of the CMS experiment during the LHC Run 1 --obtained in about 500 different
measurements carried out in \pp, \pPb\ and \PbPb\ collisions at c.m. energies
$\sqrts$~=~2.76--8~TeV-- have been summarized. They can be succinctly categorized under the following topics: 
\begin{description}
\item {\bf Quantum Chromodynamics}: In the hard sector, jet distributions are found to be in excellent 
  agreement with NLO pQCD over 14 decades in cross sections, and the strong coupling has been
  determined up to so-far unprobed scales $Q\approx$~2~TeV. Upcoming NNLO jet calculations 
  will highly profit from such measurements to better constrain the proton PDFs and $\alphas$. 
  In the semihard sector, the prominent role of multiparton interactions has been confirmed, but no
  ``beyond DGLAP'' (BFKL, saturation) QCD radiation has been observed. The bulk hadron production
  properties have helped improve the MCs for high-energy cosmic-rays physics.
\item {\bf Quark Gluon Plasma}: Signs of parton collectivity (``ridge''-like angular correlations) have been
  observed in ``central'' collisions of small systems (\pp, \pPb) with high  particle multiplicities. In
  central \PbPb\ collisions, the yields of weakly-interacting probes ($\gamma, W, Z$) are found to be
  unaffected by the QGP (and have helped constrain the nuclear PDFs), but those of strongly-interacting
  particles (jets, $b$-jets, quarkonia, high-$\pT$ hadrons) are found to be largely suppressed due to
  final-state interactions in the produced hot and dense QCD medium.
\item {\bf Electroweak}: The high-statistics differential distributions of $W$ and $Z$ bosons have improved our
  knowledge of the quark densities in the proton, while many other electroweak cross sections,
  down to the hundreds-of-fb scale, have been found in excellent agreement with (N)NLO predictions.
  First-ever electroweak measurements include: $W+t$, $\ttbar$+$\gamma$, $\ttbar+Z$, VBF production of
  the $Z$ boson, and $\gaga\to W\,W$. Multiboson processes have imposed stringent limits on anomalous triple and
  quartic gauge couplings.
\item {\bf Top quark}: Top-pair cross sections agree well with NNLO pQCD predictions and furnish a new
  competitive extraction of $\alphas$.
  A very precise measurement of the top mass, a key SM parameter chiefly connected to the
  electroweak vacuum stability, has been obtained by combining  many different $\ttbar$ final states:
  $\mtop$~=~172.38~$\pm$~0.67~GeV (0.4\% uncertainty). Single-top cross sections provide also novel
  independent constraints on the $|V_{\rm tb}|$ CKM element.
\item {\bf Higgs boson}: The last missing piece of the SM, the scalar BEH boson, was observed in 2012 in the
  high-resolution $\gaga$ and $ZZ^\star(4\Plepton)$ decay channels. Its mass, $\mH$~=~125.09~$\pm$~0.21~$\pm$~0.11~GeV,
  %obtained combining the respective CMS+ATLAS channels, 
  is known with an impressive uncertainty below
  0.2\%. Its width has been constrained through the ratio of the on- and off-shell $ZZ^\star(4\Plepton)$ decays,
  and found to be smaller than 5.4 times the SM prediction: $\Gamma_{_{\rm H}}<$~22~MeV (95\% CL).
  Its quantum numbers ($J^{PC}=0^{++}$), determined mostly through the kinematical distributions of the $ZZ^\star$
  decay mode, are those expected for a SM Higgs boson. The Higgs couplings to the $W,Z$ bosons as well as
  to the fermions ($\tau$, $b$ quark, and indirectly $t$ quark) are found to be proportional to their masses 
  (squared for $W,Z$) as expected.
\item {\bf Flavour}: The very-rare  $B^0_s\to\mu^+\mu^-$ decay with SM branching ratio BR~$\approx 10^{-8.5}$,
 considered as a ``golden channel'' for searches of SM deviations thanks to its sensitivity to virtual
 contributions from new heavy particles, has been observed with the expected BR with a statistical significance of
 6.2$\sigma$, imposing novel flavour-changing constraints in the parameter space of models beyond the SM.
\item {\bf Supersymmetry}: All searches of spartners in \pp\ final-states with excesses of $\MET$,
  same-sign leptons, multi-jets,$\gamma$,... %--often assuming large $\MET$ from the LSP in R-parity conserving
  %SUSY-- 
  have been unsuccessful so far. Such searches, mostly interpreted in terms of minimal SUSY
  implementations with a few  parameters, %(cMSSM, mSUGRA), 
  push the spartner masses increasingly away from the
  electroweak scale, and render the theory less and less ``natural'' (i.e. relevant for the resolution of  the
  SM fine-tuning problem at least).
\item {\bf Dark matter}: Mono-jet,photon searches at the LHC provide the best limits on DM searches at low masses
  ($\mDM\lesssim$~10~GeV) by exploiting boosts present in the annihilation of two partons into a DM pair.
  The derived nucleon-DM interaction cross sections limits, 
  $\sigma(\chi$-N)~$\lesssim 10^{-40},10^{-41}$~cm$^{2}$ for spin-independent and spin-dependent interactions
  respectively, %) and $10^{-41}$~cm$^{2}$ (for   spin-dependent ones),   
   cover a range of DM masses not accessible via nuclear recoils in direct underground searches.
\item {\bf Other beyond the SM searches}: No evidence of new resonances or particles connected to new
  symmetries at the TeV scale has been observed so far by looking for excesses over the SM mostly in the high invariant-mass tails
  of the distributions of pairs of jets, leptons, photons,... Stringent limits on new-physics scenarios have been imposed:
  $\Lambda\gtrsim$~15~TeV for quark compositeness (contact interactions); $\Lambda\gtrsim$~5--7~TeV for ADD gravitons;
  m$_X\gtrsim$~1.5--3.5~TeV for $W',Z'$ bosons; $\Lambda\gtrsim$~2.5~TeV for RS extra-dimensions; and
  m$_X\gtrsim$~0.6~TeV for leptoquarks, new long-lived particles, or heavy-quark partners,...
\end{description}

The lack of evidence for deviations in the data from the SM expectations is puzzling given that, apart from
the Higgs boson confirmation, the fundamental physics problems that motivated the construction of the LHC
(listed in the Introduction) remain still unsolved today. The upcoming Run-2 of the LHC, with collisions at
center-of-masses reaching the nominal 14-TeV and hundreds of fb$^{-1}$ integrated luminosities, will bring us
back to discovery mode  and hopefully to the direct (or indirect, via precision tests) observation of new
particles/symmetries at the TeV scale. %in the next years.

\paragraph*{Acknowledgments} I want to express my gratitude to the Bormio'15 organizers, the oldest nuclear \&
particle physics conference existing today, for their kind invitation to such a stimulating meeting. 

%%%%%%%%%%%%%%%%%%%%%%%%%%%%%%%%%%%%%%%%%%%%%%%%%%%%%%%%%%%%%%%%%%%%%%%%%%%%%%%%%%%%%%%%
%\section*{References}

\end{document}